\documentclass[a4paper]{jpconf}
\usepackage{graphicx}
\usepackage{xfrac}
\usepackage{nicefrac}
\usepackage{color}

\usepackage[T1]{fontenc}
\usepackage{amsthm}

\begin{document}
\title{\noindent High-pressure specific heat technique to uncover novel states of quantum matter}

\author{Julio Larrea J. $^{1}$, V. Martelli $^{1}$, H. M. R\o{}nnow $^{2}$}

\vspace{0.2cm}

\address{$^{1}$ Institute of Physics, University of S\~{a}o Paulo, 05508-090 S\~{a}o Paulo, Brazil}

\vspace{0.2cm}




\address{$^{2}$ Laboratory for Quantum Magnetism, Institute of Physics, \'{E}cole Polytechnique F\'{e}d\'{e}rale de Lausanne, Switzerland}





\ead{larrea@if.usp.br}

\begin{abstract}

AC-specific heat measurements remain as the foremost thermodynamic experimental method to underpin phase transitions in tiny samples. However, its performance under combined extreme conditions of high-pressure, very low temperature and intense magnetic fields needs to  be broadly extended for investigation of quantum phase transition in strongly correlated electron systems. In this communication, we discuss the determination of specific heat on the quantum paramagnetic$-$insulator SrCu$_{2}$(BO$_{3}$)$_{2}$ by applying the AC-specific heat technique under extreme conditions. In order to apply this technique to insulating samples we sputtered a metallic thin film-heater and attached thermometer onto sample. Besides that, we performed full frequency scans with the aim to get quantitative specific heat data. Our results show that we can determine the sample heat capacity within 5$\%$ of accuracy respect to an adiabatic technique. This allows to uncover low energy scales that characterize the ground state of quantum spin entanglement in SrCu$_{2}$(BO$_{3}$)$_{2}$.

\end{abstract}

\section{Introduction}

In the last two decades, the condensed matter physics community has increased the attention towards the use of the temperature-modulation calorimetry (AC calorimetry) to investigate the ground state of strongly correlated electron systems ~\cite{Wilhelm_2002, Lortz2005, Park2006, Knebel2011, Larrea2016, Poran2017, Shen2020}. The current challenge has focused to adapt this experimental method to be used under the application of high-pressure as a control parameter and to be capable to track with reliable accuracy the evolution of a phase transition across a quantum critical point (QCP), which is the phase transition that separates two states of matter at $T=$ 0 K. Following the change of the temperature dependence of the sample heat capacity, $C_{S}(T)$, one can distinguish either the suppression of the phase transition or the emergence of a new type of collective quantum phenomena at the QCP ~\cite{Wilhelm_2002, Lortz2005, Park2006, Knebel2011, Larrea2016, Poran2017, Shen2020}. 

In the investigation of heavy fermion (HF) systems, AC calorimetry has been used as a relevant tool to elucidate complicated pressure ($p$) $-$ temperature ($T$) phase diagrams highlighting a better understanding about the thermodynamic behavior of the physical degrees of freedoms. For instance, experiments show that the anomaly in $C_{S}(T)$ associated with the magnetic order can be suppressed by pressure ($P$) continuously (or discontinuously) towards $T = 0$ K, sometimes with the emergence of unconventional superconductivity either in bulk ~\cite{Wilhelm_2002, Lortz2005, Park2006, Knebel2011, Shen2020} or in thin film samples ~\cite{Poran2017}. Besides that, an accurate measurement of $C_{S}(T)$ under high pressures $-$ which is a very challenging experiment $-$ allows to infer relevant information about the fate of low-lying excitations in the vicinity of a QCP associated with novel quantum states of matter both with trivial and non-trivial topology ~\cite{Lai2017, Wietek2019}. However, such accurate measurements under high pressure continue to be scarce in comparison to those $C_{S}(T)$ measurements performed at ambient pressure ($p=$ 0) and under adiabatic conditions, the latter reaches an inaccuracy $<$ 1$\%$ ~\cite{Gmelin1997}.    

\begin{figure}[th]
	\centering \includegraphics[angle=0,scale=0.6]{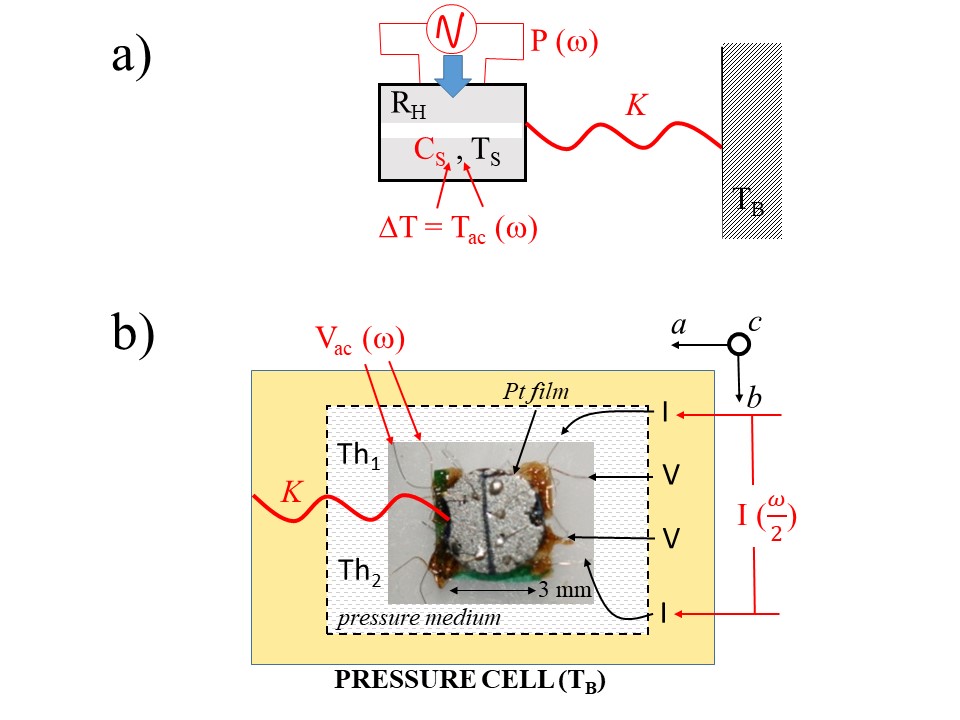}
	\caption{(Color online) Schematic arrangement of the relevant parameters involved in the steady-state equation for our current AC calorimetry experiments under $p$ and $T$ in a second harmonic (2$\omega$) mode. (a) Representation as block diagram  (b) Photo of the slab sample SrCu$_{2}$(BO$_{3}$)$_{2}$ capped with Pt thin film of 20 nm thickness deposited onto the surface that lies within the $a$-$b$ axis. This assemble is glued with GE varnish onto a VESPEL platform. The dashed-rectangle line represents the teflon cup of the pressure cell. The arrows show the leads associated with two thermocouple wires ($Th_{1}$ and $Th_{2}$) and with 4-point electrical resistance measurement for the heater (I$-$V$-$V$-$I). $P(\omega)$ $-$ the AC oscillating heat flow applied through the leads I$-$I; $K$ $-$ thermal link with conductance between sample and pressure cell setup; Temperatures of $T_{S}$ $-$ sample; $T_{ac} (\omega)$ oscillating sample (measured as $V_{ac} (\omega)$, the oscillating pick-voltage signal in $\omega$); $T_{B}$ $-$ thermal bath or temperature of the pressure cell body; $C_{S}$ $-$ sample heat capacity.}%
	\label{fig1}%
\end{figure}

Unlike the case of heavy fermion systems, the AC calorimetry under high pressure has seen much less application in the investigation of low-dimensional frustrated quantum magnets. For the latter, it is due either to the scarcity of synthesized samples or because these systems grew mainly as almost perfect insulators with intrinsic very low sample thermal diffusivity $\eta$ even at low $T$. A low $\eta$ implies a large internal time relaxation ($t_{in}$) to achieve thermal equilibrium. It drives the working frequency ($\omega_{c}$), i.e. $-$ the characteristic frequency to maximize the sample heat capacity through the measurement of the oscillating sample temperature ($T_{ac}$)$-$, to reach very small values (typically $\omega_{c}$ is of the order of 1 Hz in insulators) \cite{Baloga1977}. This working frequency is, at least, an order of magnitude lower than it is found for metals \cite{Wilhelm_2002, Lortz2005, Sullivan1968, Eichler1979}. Remarkably, very low $\omega_{c}$ might become a problem for the measurement duration because the smaller is $\omega_{c}$ the longer has to be the acquisition-data time to pick up the oscillating voltage signal $V_{ac}$ (which is proportional to $T_{ac}$). In addition, another challenging problem for insulators inside a high-pressure cell is the control of the thermal link conductance ($K$) between sample (and thus its sample temperature $T_{S}$) and other parameters relevant in the steady-heat equation \cite{Sullivan1968, Eichler1979, Gmelin1997} such as the heater ($R_{H}$), the temperature of the body cell (or bath temperature $T_{B}$), the pressure medium, the excitation heat flow ($P(\omega)$) and temperature ($T_{ac}( \omega)$) (see figure \ref{fig1}a).

Here, we present an alternative route to obtain the accurate determination of the temperature $-$ dependence of sample heat capacity under high pressure. We present and discuss an experimental setup of $\textit{lab on a slab sample}$, which consists of metallic thin films deposited onto the insulator surface SrCu$_{2}$(BO$_{3}$)$_{2}$ with the aim to optimize the sample oscillating temperature associated with $C_{S}$ as well as to have a better control of the relevant parameters involved in the steady-state equation. Finally, we propose a methodology to obtain absolute values of $C_{S}(T)$ and the estimation of the thermal link between sample and surrounding, which are computed from state-of-the-art data analysis of isothermal frequency scan fits. Our results show that we can guarantee the determination of $C_{S}(T)$ under high pressure, at least, within 5 $\%$ of accuracy respect to the adiabatic technique ~\cite{Gmelin1997}.

\section{Experimental}

SrCu$_{2}$(BO$_{3}$)$_{2}$ single crystal samples were grown by the traveling solvent floating zone technique with a method similar to Ref.~\cite{Gaulin2004}, having the same quality of batch sample as reported by Zayed $\textit{et. al.}$ ~\cite{Zayed2017}. 

The temperature dependence of the heat capacity, $C(T)$, was measured using the alternate current (AC) calorimetry at a second harmonic (2$-$$\omega$) mode ~\cite{Baloga1977, Sullivan1968, Eichler1979, Gmelin1997}. Samples were cut as a slab with a cross section of average side $\approx$ 3 mm within the $ab$ plane and thickness parallel to the $c$-axis of $t \approx$ 0.6 mm (see figure \ref{fig1}b). Pt thin films of 20 nm thickness (grey regions) were deposited onto the surface as shown in figure \ref{fig1}b. The whole assemble, i.e., sample and Pt thin film, was glued with GE varnish to a VESPEL sample holder. The separation of the Pt thin films in two parts allows to glue two different thermocouples ($Th_{1}$ and $Th_{2}$) and a heater, measuring for the latter its $T$-dependence of the electrical resistance $R_{Pt}(T)$ in a separate measurement by a standard four-probe measurement. As thermocouple we used a pair of AuFe(0.07)/ Chromel wires of 25 $\mu$m diameter. The idea to mount two different thermocouples was to infer some possible thermal gradient along the sample. However, a sizable thermal gradient was not detected within the resolution of our AC-calorimetry measurement. On the hand, constantan wires of 25 $\mu$m diameter serve as leads to conduct an oscillating current of intensity $I_{0} =$ 3.5 mA through the Pt thin film heater. All leads were glued onto the sample using a silver epoxy H31LV that ensures a good thermal conductivity.

To load the pressure, we used a piston cylinder BeCu clamp cell, which presents 27 kbar as the upper limit of pressure. The pressure transmitting medium was liquid kerosene and a Pb strip was used as pressure manometer at low temperatures. This pressure cell was inserted into a cryostat that allows to cover temperatures down to 3 K. At each pressure step, we observed good hydrostatic conditions around the sample with reproducible results in different batches of sample.
 
Frequency  scans ($f$-scan) at fixed temperature (isothermal) and at constant pressure (isobar) were performed in order to investigate the route to get a more appropriate working frequency ($\omega_{C} =$ 2 $\pi f_{C}$) to measure accuracy $C_{S}(T)$. The experimental $f$-scans demanded a  very careful data acquisition and temperature stabilization. Because these $f$-scans were performed from frequencies lower than 1 Hz, we needed to set long time constant to acquire lock-in voltage at each frequency. Moreover, the stability of the bath temperature ($T_{B}$) was controlled to vary within a temperature range narrow enough to avoid spurious contribution in the voltage signals of the thermocouple thermopower AuFe(0.07)/ Chromel, $S_{AuFe} (T)$, even at the investigated high-pressures  ~\cite{Choi2002}. 


\section{Results and Discussion}

The relevant parameters in our AC calorimetry setup are shown in figure \ref{fig1}a. An oscillating current with amplitude $I_{0}$, i.e, $I(\frac{\omega}{2}) =$ $I_{0} e^{i\frac{\omega}{2}t}$ is applied on a Pt thin film heater that has an electrical resistance of $R_{H}$. The oscillating heat flow $P(\omega) =$ $P_{0}$ $e^{i \omega t}$, where $P_{0} = I_{0}^{2} R_{H}$, propagates into the sample with a frequency $\omega = 2 \pi f$, being $f$ the frequency set in the lock-in. Within this arrangement, the sample heat capacity ($C_{S}$) can be derived from the steady$-$state heat equation ~\cite{Baloga1977, Sullivan1968, Gmelin1997}: 

\begin{equation}
T_{ac} = \frac{P_0}{K + i \omega C_S}  
\end{equation}

with $i^{2} = -$1, $T_{ac}$ the oscillating sample temperature and $K$ the conductance of the thermal link  between sample and the surrounding of the pressure cell (which includes wires, the sample holder, the pressure medium, the teflon cap and the body cell). Equation (1) can be also written using the oscillating pick-up voltage ($V_{ac}$) and our thermocouple thermopower sensitivity $S_{AuFe}$: 

\begin{equation}
V_{ac} = \frac{1}{K^{'} + i f C_S^{'}} 
\end{equation}

where  $K^{'} = \nicefrac{K}{(S_{AuFe} P_0)}$ and $C_S^{'} = \nicefrac{2 \pi C_S}{(S_{AuFe} P_0)}$. Fitting the $f$-dependence of $V_{ac}$ with equation (2) we computed directly the quantities $K^{'}$ and $C_S^{'}$, the latter related to the sample heat capacity.

\begin{figure}[th]
	\centering \includegraphics[angle=0,scale=0.45]{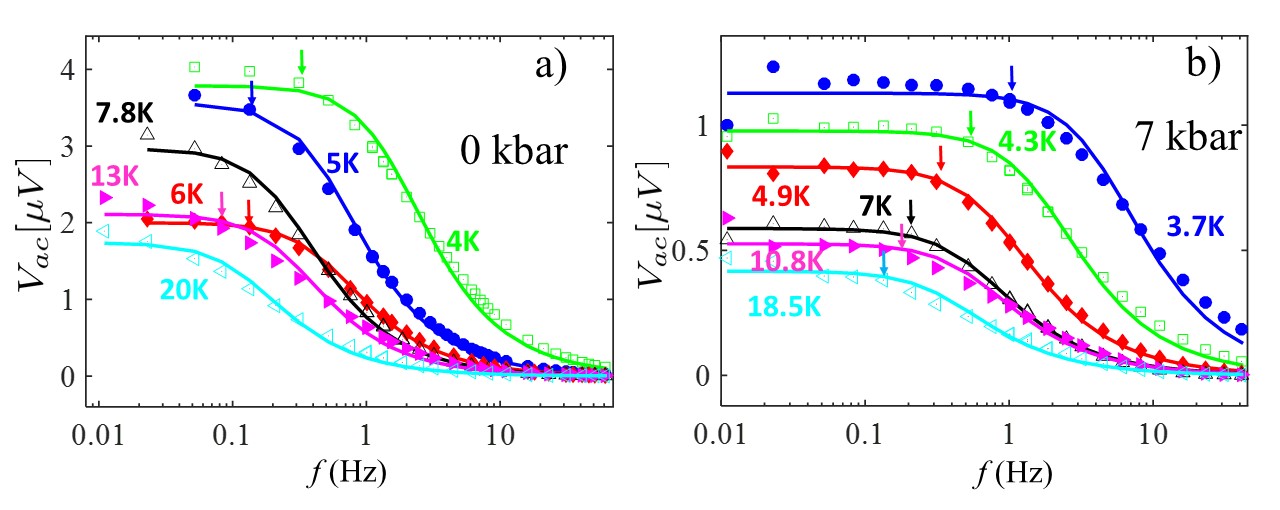}
	\caption{(Color online) Representative isothermal measurements of the frequency ($f$)-dependence of the oscillating pick-up voltage ($V_{ac}$) on  SrCu$_{2}$(BO$_{3}$)$_{2}$ single crystal and pressure: (a) Ambient pressure ($p_{1} =$ 0 kbar)  (b) high pressure $p_{2} =$ 7 kbar. The solid lines represent fits using equation (2).}%
	\label{fig2}%
\end{figure}

Figure \ref{fig2} shows the isothermal $f$-dependence of $V_{ac}$ at two different pressures $p_{1} =$ 0 kbar $p_{2} =$ 7 kbar. We can observe in figure \ref{fig2} that at the highest frequencies $V_{ac}$ tends to vanish revealing a very small $t_{in}$ associated with a heat capacity of samples of very small volume. Decreasing the frequency, we observe that $V_{ac}$ increases reaching a plateau at the cutoff frequency, $f_{off}$, (see downwards arrows in Figure \ref{fig2}), which reaches higher values for $p_{2} =$ 7 kbar. For both pressures, at frequencies below $f_{off}$, $V_{ac}$ saturates to constant values indicating another time scale, $\tau_{off} \propto 1 / f_{off}$, for which the heat flows through the sample and surrounding. This is consistent with the relationship between the Joule heating contribution and the DC offset pick-up voltage, $V_{dc}$, at $f =$ 0, i.e., $V_{dc} = \nicefrac{(S_{AuFe} P_0)}{K}$.

\begin{figure}[th]
	\centering \includegraphics[angle=0,scale=0.15]{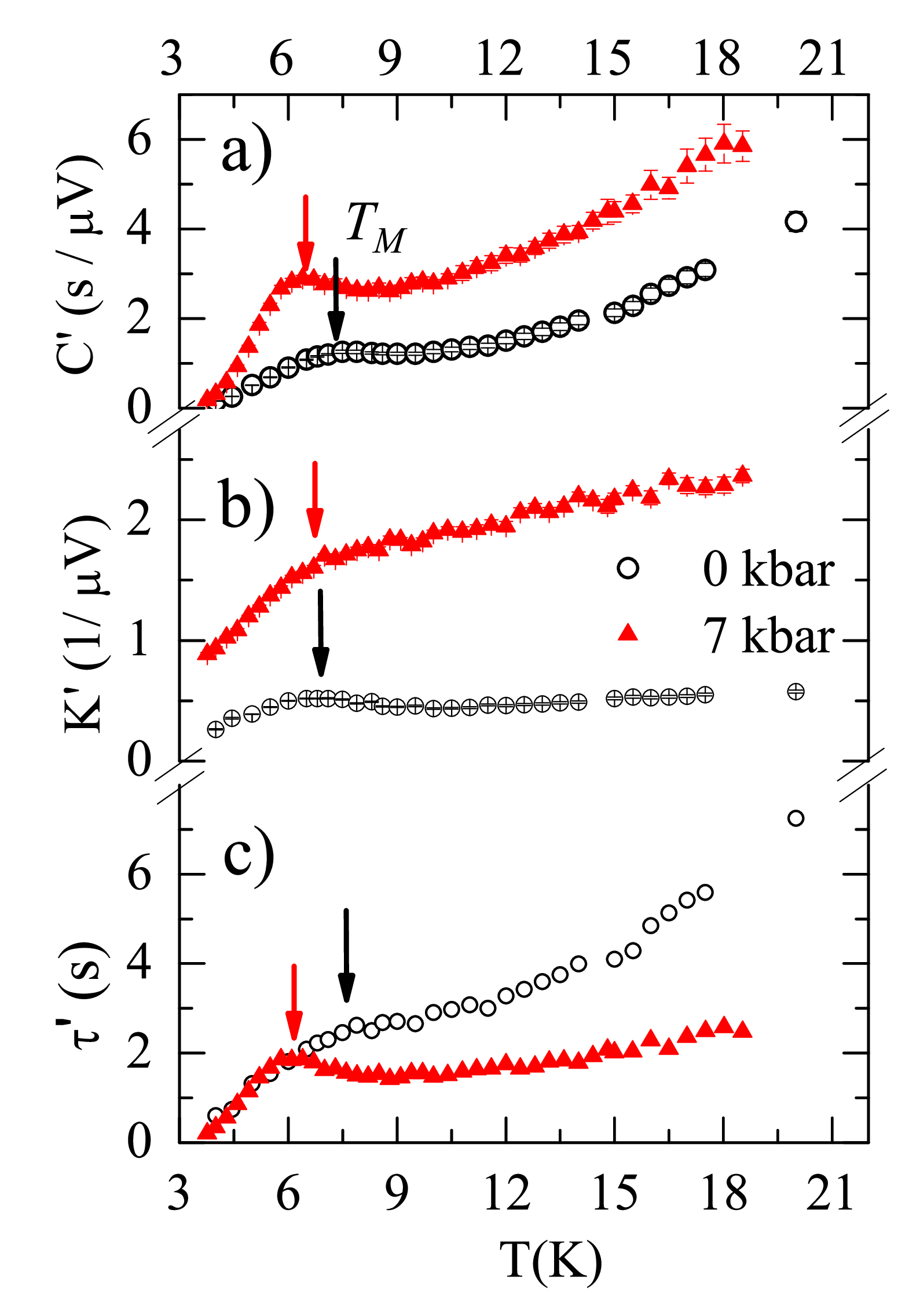}
	\caption{(Color online) Temperature dependence of the parameters obtained from the fit of equation (2) at different pressures: (a) Quantity proportional to the sample heat capacity $C_S^{'}$  (b) Quantity proportional to conductance of thermal contact between sample and surrounding $K^{'}$. (c) The time scale $\tau^{'} = C_S^{'} / K^{'}$. Error bars  are also included in the calculated parameters obtained by the fitting algorithm.}%
	\label{fig3}%
\end{figure}

It has been assumed for a sample with metallic behavior that the working frequency ($\omega_{C}$) should be selected in the range $ 1 / \tau_{off} < \omega_{C} < 1 / t_{in}$ \cite{Wilhelm_2002, Lortz2005, Sullivan1968, Eichler1979}. The choice of this $\omega_{C}$ is not trivial in the case of an insulator due to the narrow window of frequency range originated from the sample low thermal diffusivity and the insulating pressure medium. In addition, the application of high pressure might influence the characteristic time scales and thus $\omega_{C}$, which for instance at 0 kbar, it could reach values lower than 1 Hz. All these claims are clearly shown in Figure \ref{fig2} where the plateau in $V_{ac}(f)$ at $p_{1} =$ 0 kbar is not well defined as it seen $p_{2} =$ 7 kbar. 

A first attempt criteria for the choice of $\omega_{C}$ is to satisfy $\omega_{C} \tau >>$ 1, being $\tau = \nicefrac{C_S}{K}$, a characteristic time scale to maximize the sample heat capacity as $C_{S}= \nicefrac{(S_{AuFe} P_0)}{\omega_{C} V_{ac}}$. Within the limitation to determine precisely $\omega_{C}$, $C_{S} \propto \nicefrac{1}{V_{ac}}$ measured at a frequency close to $\omega_{C}$ is quite often enough to resolve phase transitions in $T$ $-$ dependence of sample heat capacity but not to obtain measurements of absolute values of $C_{S}$  ~\cite{Wilhelm_2002, Lortz2005, Park2006, Knebel2011, Larrea2016}. 

In order to get advanced with accurate measurements of $C_{S} (T)$ under pressure, we propose the determination of $\omega_{C}$ from the fit of $f$-scan data analysis of figure \ref{fig2}. Solid lines show the fits obtained by using equation (2). The relevant parameters obtained from the fits, $C_S^{'}$, $K^{'}$ and the ratio $\tau^{'} = C_S^{'} / K^{'}$ (being $\tau^{'} = 2 \pi \tau$), at different pressures, are plotted in figure \ref{fig3}. The expression in equation (2) describes quite well the experimental data overall $f$-scan, mainly at 7 kbar, where the cutoff frequency, $f_{off}$, sets at higher values. The validity of the model in equation (2) reveals that our experimental setup shown in figure \ref{fig1} keeps the condition of good thermal conductance between sample, thermocouple and heater.

\begin{figure}[th]
	\centering \includegraphics[angle=0,scale=0.15]{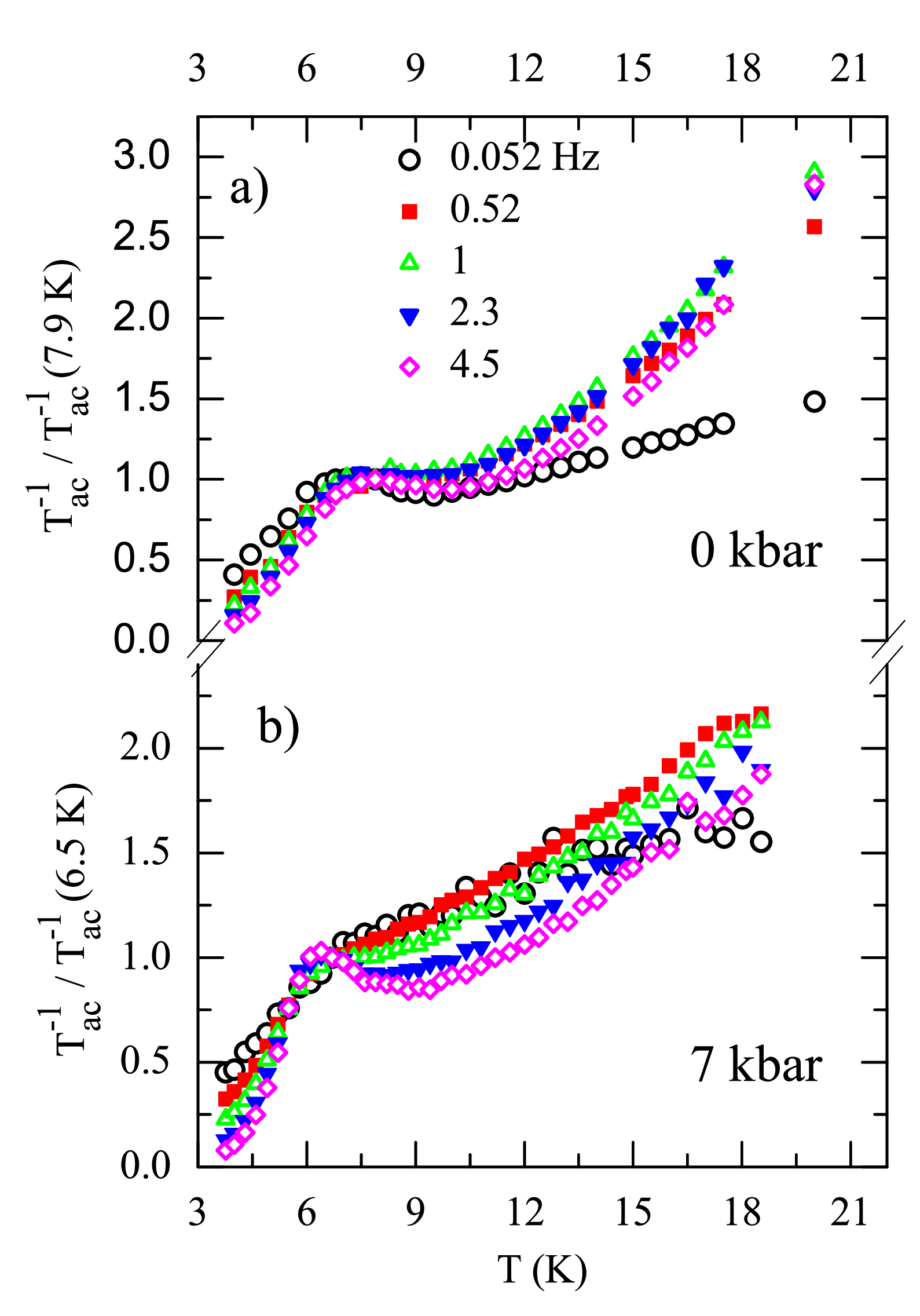}
	\caption{(Color online) Temperature scan of the inverse of the oscillating sample temperature ($T^{-1}_{ac}  \propto C_{S}$) at fixed frequencies for different pressure: (a) 0 kbar and  (b) 7 kbar.  Values of $T^{-1}_{ac}$ are normalized for those temperatures at $T_{M}$ (see fig. \ref{fig3}a). }%
	\label{fig4}%
\end{figure}

Figure \ref{fig3}a shows the $T$-dependence of a quantity proportional to the sample heat capacity $C_S^{'} (T)$. For both pressures we can clearly observed anomaly in $C_S^{'} (T)$ whose maximum occurs at temperatures ($T_{M}$) (see also downwards arrows). $T_{M}$ decreases with pressure, following the same decreasing rate ($T_{M} (7 kbar)$ / $T_{M} (0 kbar) \approx$  0.8) as it was reported for the $p$-dependence of the spin singlet-triplet gap ($\Delta (p)$) in SrCu$_{2}$(BO$_{3}$)$_{2}$ measured by inelastic neutron scattering ~\cite{Zayed2017}. This coincidence supports the reliability of $f$-scan data analysis in eq. (2) to resolve anomaly position in heat capacity measurements. 

\begin{figure}[th]
	\centering \includegraphics[angle=0,scale=0.15]{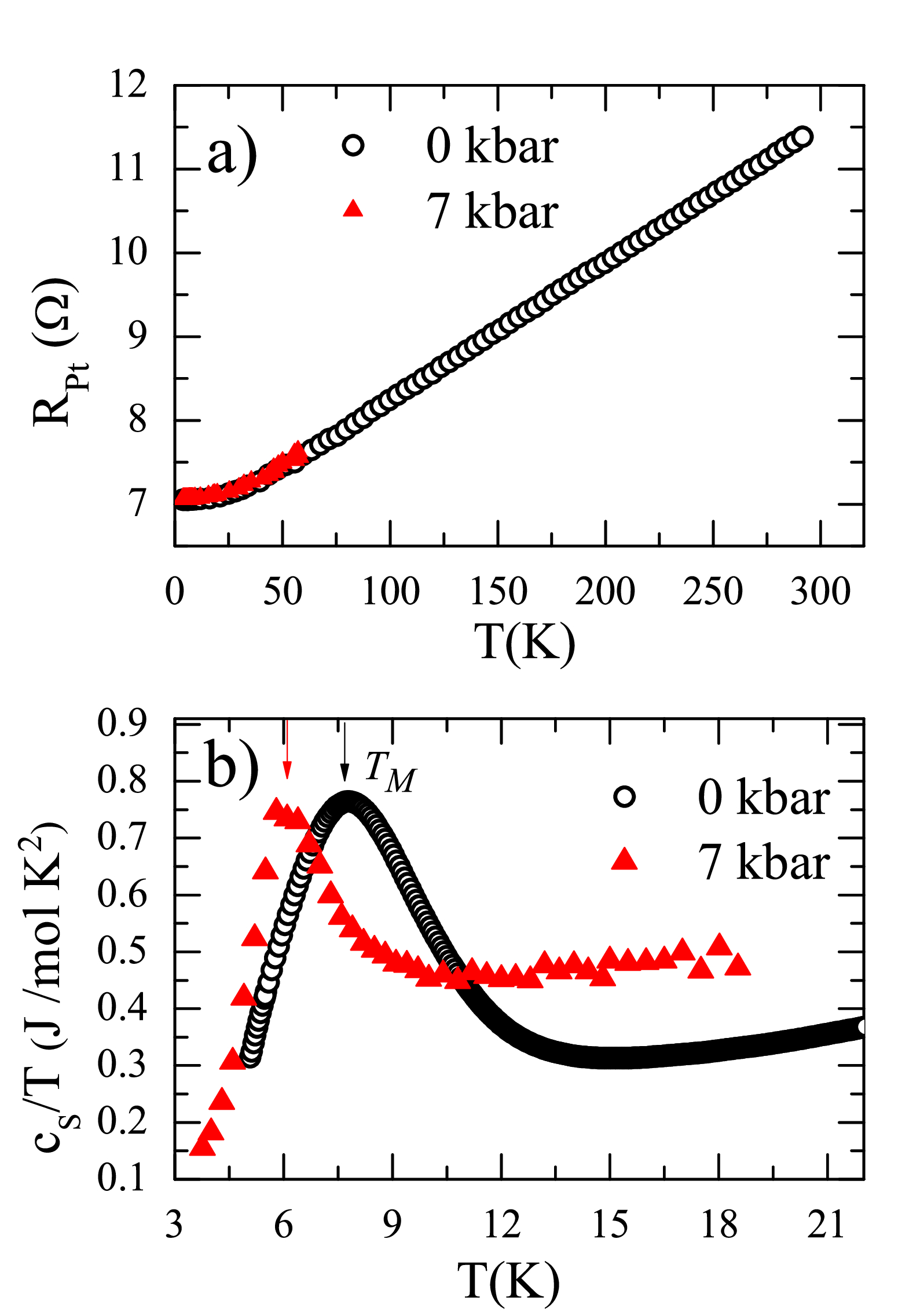}
	\caption{(Color online) Temperature dependence of physical property measurements at different pressures presented in equation (3) (a) Electrical resistance of the Pt film with 20 nm thickness (b)  Absolute values of the sample specific heat over temperature of SrCu$_{2}$(BO$_{3}$)$_{2}$. Downwards arrows show the temperatures where the curves have a maximum at $T_{M}$}%
	\label{fig5}%
\end{figure}

On the other hand, Figure \ref{fig3}b depicts that $K^{'}(T)$ changes the slope at temperature around $T_{M}$, but its position does not change with pressure. This is an indication that $K^{'}(T)$ involves mainly contribution from the surrounding rather than the sample. For the current experiment, it is very difficult to separate the contributions in $K^{'}(T)$ because the thermal conductance between the surrounding and sample includes the relevant parameters in the steady state equation of our AC-calorimeter together with the pressure medium and pressure cell accessories depicted in \ref{fig1}b. Nevertheless, all together these contributions determine the plateau observed below $f_{off}$ (downwards arrows in fig. \ref{fig2}). Instead, a better interpretation of $K^{'}(T)$ can be done from the ratio $\tau^{'} (T) = \nicefrac{C_S^{'} (T)}{K^{'}(T)}$ plotted in figure \ref{fig3}c. We observe that decreasing the temperature, $\tau^{'}$ changes the slope at the same temperature where $C_S^{'} (T)$ does. At such inflection points (see downwards arrow), we can estimate a relative decreasing of $\tau$ as $ \Delta \tau / \tau = \tau^{'} (p_{1}) - \tau^{'}(p_{2}) / \tau^{'} (p_{1}) \approx$ 25 $\%$. The time scale $\tau$ sets the condition of the critical frequency $\omega_{C}$ (i.e., $\omega_{C} \tau >>$ 1) for which the sample heat capacity maximizes. Therefore, figure \ref{fig3}c shows that the working frequency $\omega_{C}$ increases with pressure, at least, within the investigated range of pressure.

In order to infer more accuracy determination of $\omega_{C}$ we performed isobar $T$-scan at different fixed frequencies. The selected frequencies are chosen from fig. \ref{fig2} which depicts that $\omega_{C}$ might be at frequencies around or lower than 1 Hz and 4 Hz, for $p_{1}=$ 0 kbar and $p_{2}=$ 7 kbar, respectively.  Figure \ref{fig4} shows the inverse of the oscillating temperature ($T^{-1}_{ac} $) normalized at the temperature $T_{M}$, the latter estimated from fig. \ref{fig3}a. The value of $T^{-1}_{ac} $, which is proportional to the sample heat capacity at $\omega_{C}$, was computed from the experimental pick-up voltage as $T_{ac} = V_{ac}/ S_{AuFe}$. In principle, it is expected that the distancing from the expected $\omega_{C}$ causes a broadening and shifting of the anomaly position observed in $C(T)$. Thus, an optimal choice of $\omega_{C}$ should reveal the narrowest anomaly and maximum absolute values of $C(T)$ together with most accuracy determination of the anomaly position. Knowing the anomaly position obtained from our $f$-scan (see fig. \ref{fig3}a), our inspection of fig. \ref{fig4}, reveals that these requirements are satisfied for working frequencies, $\omega_{C} / 2 \pi$, at 0.052 Hz and 2.3 Hz, for 0 kbar and 7 kbar respectively.

\begin{figure}[th]
	\centering \includegraphics[angle=0,scale=0.15]{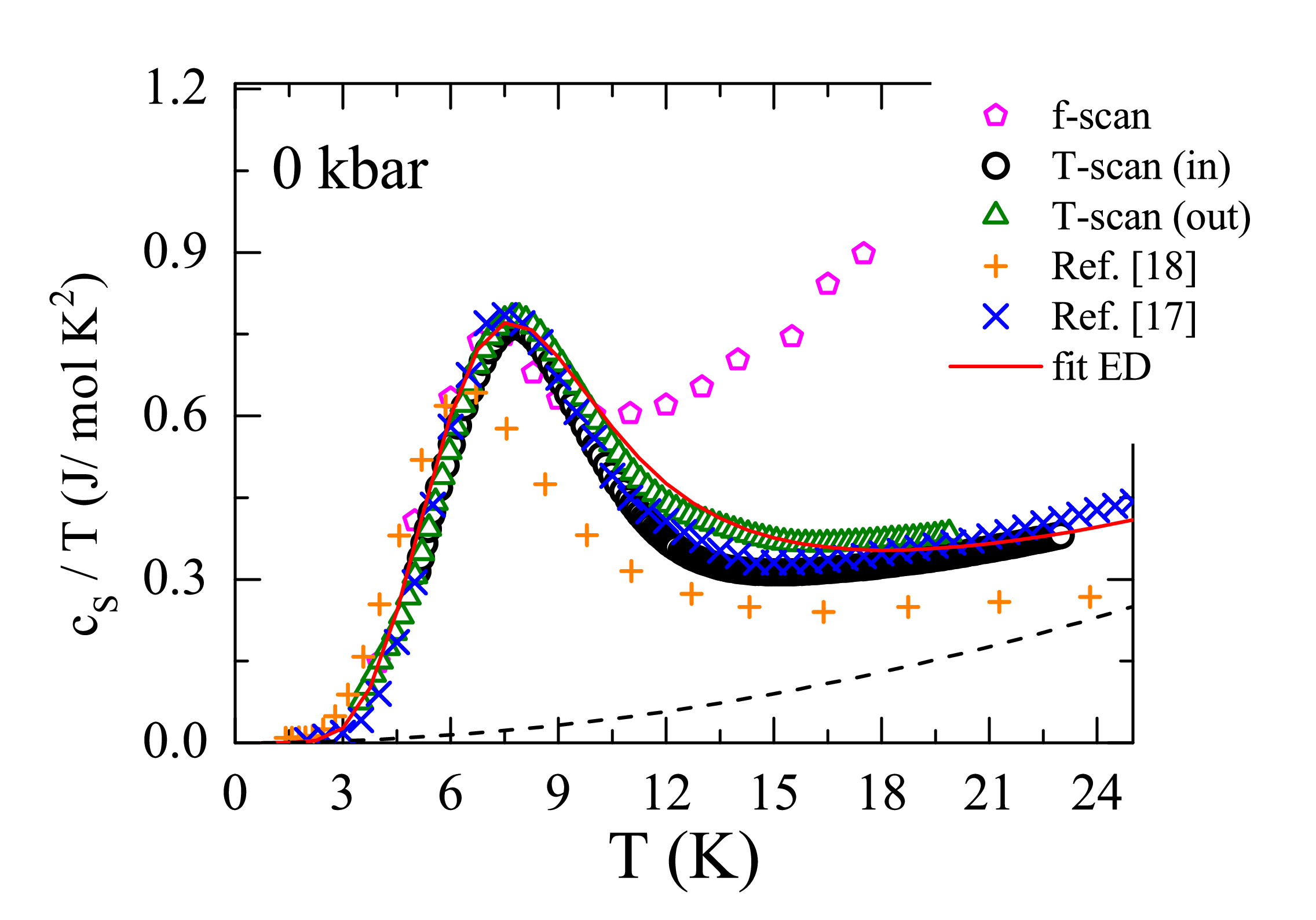}
	\caption{(Color online) Sample specific heat over temperature of SrCu$_{2}$(BO$_{3}$)$_{2}$ measured from $f$-scan and $T$-scan with a fixed frequency $=$ 0.052 Hz, inside (in) and outside (out) the pressure cell. For $T$-scan measurement, data is multiplied by a factor of 0.15 (see text). For comparison we plot data reported in Ref. ~\cite{Kageyama2000} and Ref. ~\cite{Jorge2005} at adiabatic conditions. The solid line is the fit considering both the diagonalization (ED) and the phonon contribution. Dashed line is the phonon contribution as $c_{ph} = \beta \times T^{3}$ with $\beta =$ 4$\times$10$^{-4}$ J/mol K$^{4}$ (see text).}%
	\label{fig6}%
\end{figure}
 
After the selection of the working frequency, we can evaluate the absolute values of the overall $T$-dependence of our sample specific heat ($c_{s}$) by using the following relation:

\begin{equation}
c_{s} = \frac{ I_{0}^{2} P_{A} R_{Pt} } {M \omega_{C} T_{ac} }  
\end{equation}

where $P_{A} =$ 332.334 gr/ mol is the molecular weight for SrCu$_{2}$(BO$_{3}$)$_{2}$, $M =$ 0.0152 gr the sample mass. The temperature dependence of the electrical resistance of the Pt heater, $R_{Pt} (T)$, is shown in figure \ref{fig5}a. The absolute values of the electrical resistance is consistent with a Platinum thin film with 20 nm thickness whose residual resistivity ratio $ RRR =$ 1.63 shows that the heater at present experiment has a good metallic behavior ~\cite{Berry1963, Dutta2017} and negligible pressure dependence inside the investigated pressure range ~\cite{Antonov1984}. Knowing all relevant parameters in equation (3) we computed the sample specific heat, whose quantity over temperature, $c_{S}/T$ , is plotted as function of temperature in figure \ref{fig5}b. The actual temperature, $T$, in $c_{S} (T)$ is corrected as $T= T_{B} + T_{dc}$, where $T_{B}$ is the bath temperature or temperature of the thermometer attached to the pressure cell (see fig. \ref{fig1}a) and $T_{dc}$, the DC-offset Joule heating measured separately. 

In order to evaluate absolute values of $c_{S} (T)$ for 0 kbar, we plot in figure \ref{fig6} the comparison of $c_{S} (T) / T$  between the values obtained from $f$-scan and $T$-scan analysis with those data extracted from Ref. ~\cite{Kageyama2000, Jorge2005} which were obtained at adiabatic conditions. For the $f$-scan, the sample specific heat is obtained as $c_{S}= \nicefrac{(S_{AuFe} I_{0}^{2} P_{A} R_{Pt}) C_S^{'}}{2 \pi M }$, where $C_S^{'}$ is shown in fig. \ref{fig3}a. We can see in fig. \ref{fig6} that $c_{S} (T) / T$ obtained for our $f$-scan analysis matches very well the data obtained in Ref. ~\cite{Kageyama2000}, at least, for all $T \leq$ 10 K. A possible explanation for the discrepancy at $T >$ 10 K might be attributed to the influence of the addenda contribution to the sample heat capacity which is difficult to remove it from calculated values of $C_S^{'}$ using eq. (2) and above 10 K.

In addition, figure \ref{fig6} shows the $c_{S} (T) / T$ at fixed frequency of 0.052 Hz, measured when the sample was inside ($in$) (in a liquid medium) and outside ($out$) (vacuum) the pressure cell. The values of $c_{S} (T)$ were multiplied by a factor of 0.15 to match the height of the anomaly in $c_{S} (T) / T$ obtained by $f$-scan. This factor might indicate that at such low-frequency, the amplitude of the oscillating temperature is so high that only 15 $\%$ of the sample mass is measured effectively with maximum heat capacity. With the same criteria we multiplied $c_{S} (T)$ data for 7 kbar by a factor of 0.5 to obtain absolute values of sample specific heat under pressure plotted in fig. \ref{fig5}b. The factors chosen at 0 kbar and 7 kbar are in complete agreement with the ratio between $V_{ac}$ observed at 0 kbar and 7 kbar in fig. \ref{fig2}. 

\begin{figure}[th]
	\centering \includegraphics[angle=0,scale=0.5]{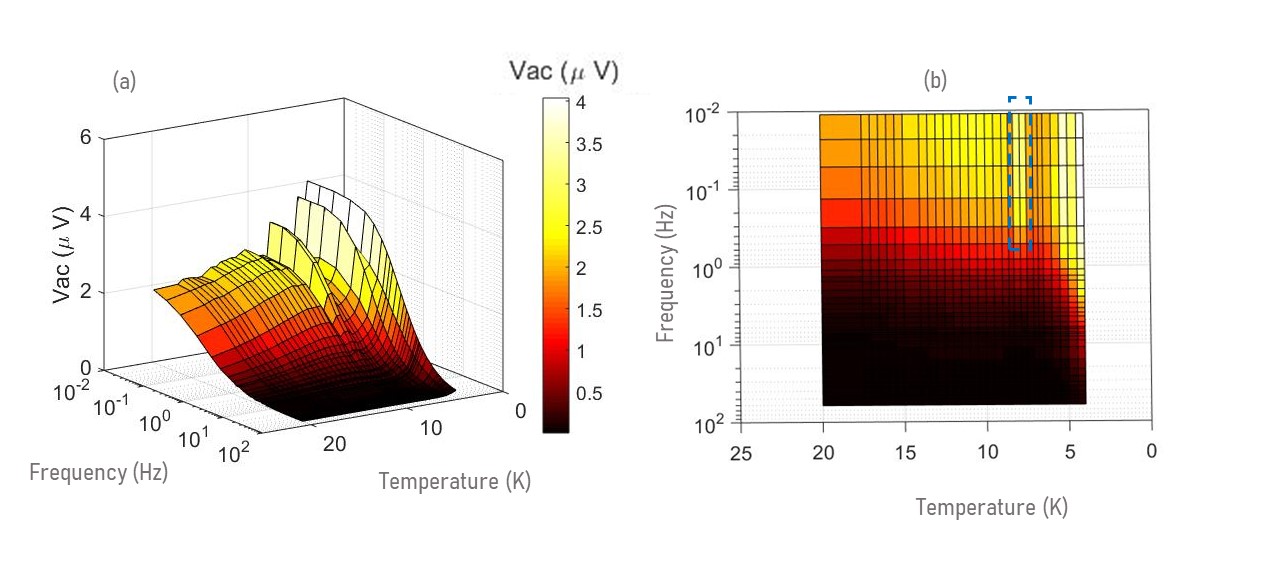}
	\caption{(Color online) Color map of the isothermal $f$-scan data at $p_{1} =$ 0 kbar of SrCu$_{2}$(BO$_{3}$)$_{2}$. (a) A three-dimensional temperature$-$frequency$-$ oscillating pick-up voltage ($V_{ac}$) plot, whose module for the latter is also shown on  right side. (b) The projection on the temperature$-$frequency axes. The dashed rectangle shows region where the sample heat capacity is maximum, within an accuracy $\sim$ 5 $\%$. The dark color shows the forbidden region of frequencies where the sample heat capacity is negligible.}%
	\label{fig7}%
\end{figure}

Remarkably, our $c_{S} (T) / T$ data for 0 kbar shows an excellent agreement with the literature data in  Ref. ~\cite{Kageyama2000} overall T-range. Such finding indicates the right choice of the working frequency $\omega_{C}/ 2 \pi =$ 0.052 Hz. Besides that, the fact that we obtained similar results for the sample specific heat at different media surrounding of the sample highlights that right choice of $\omega_{C}$ which can indeed maximize the sample heat capacity without to concern in addenda contribution.

Other indications that support our methodology to obtain absolute values of $c_{S}$ can be inferred from the analysis of $c_{S} (T)$. In particular, at ambient pressure $p_{1} =$ 0 kbar, the ground state of SrCu$_{2}$(BO$_{3}$)$_{2}$ is the realization of the square spin Shastry-Sutherland lattice ~\cite{Shastry1981}, whose solvable Hamiltonian gives an exact solution as a product of two orthogonal dimers characterized by a spin singlet-triplet gap $\Delta \approx$ 34.8 K. More recently, inelastic neutron experiments under high pressure  ~\cite{Zayed2017} revealed that this gap decreases with pressure as $\Delta (p) =$ 34.8 K $-$ 0.928 K /kbar $\times p$. Because the position of the anomaly in $c_{S} (T)$ is proportional to the gap, we can infer from fig. \ref{fig5}b that our $c_{S}$ data under pressure reproduces excellently the position of the anomaly in $c_{S} (T)$, so that $T_{M} (7 kbar) / T_{M} (0 kbar) = \Delta (7 kbar) / \Delta (0 kbar) =$ 0.81. Besides that, using the exact diagonalization calculation for $N = 20$ spins, with an inter-dimer $J^{'}$ to intra-dimer $J =$76 ratio $\alpha = J^{'} / J =$ 0.61, added to a phononic contribution $= \beta T^{3}$ (with $\beta =$ 4$\times$10$^{-4}$ J/mol K$^{4}$) we fit quite well our experimental $c_{S} (T)$ data (see solid lines in fig. \ref{fig6}), obtaining a gap $\Delta \approx$ 36 K, a value very close to what is expected for the ground state of the spin square lattice SrCu$_{2}$(BO$_{3}$)$_{2}$. 

Finally, an experimental three-dimensional color map plot that correlates temperature, frequency and the pick-up voltage signal is shown in Fig. \ref{fig7}a for $p_{1} =$ 0 kbar. A most defined shape of oscillating $V_{ac}$ signal with highest values are clearly seen at low frequencies. On the other hand, very small $V_{ac}$, registered with dark color, might be associated to a forbidden zone of frequencies where the sample heat capacity becomes undetectable. The projection at the temperature$-$frequency axis plotted in fig. \ref{fig7}b reveals more clearly this forbidden zone of frequencies. Remarkably, the working frequency for 0 kbar is located in the region marked with lighter colors, covering values of $V_{ac} \geq$ 2 $\mu$V. According to thermocouple thermopower AuFe(0.07)/ Chromel, $S_{AuFe} (T)$, this number demands to have a control in sample temperature stability with a fluctuation not higher than 0.1 K. Such order of fluctuation can be much lower at very low temperature range. Besides that, it is very clear from fig. \ref{fig7}b that the lighter region reveals that the frequency to measure the signal of the heat capacity at 0 kbar should fit inside the region lower than 1 Hz. In particular, the widest frequency region where we can measure the sample heat capacity occurs at that temperature where the anomaly in the sample heat capacity reaches its maximum (see dashed rectangle line in fig. \ref{fig7}b).

\section{Conclusions}

In conclusion, we proposed a methodology to use the AC-calorimetry to measure accurate heat capacity of an insulator at extreme conditions of high pressure and low temperature. Our proposal to sputter a metallic thin film on the sample ensures the condition of good thermal conductance between sample, heater and thermometer. This allows a reasonable determination of the absolute values of sample heat capacity by the analysis of isothermal frequency scans regarding steady$-$state heat equation. From the isothermal $f$-scan we can also infer the working frequency for which the sample heat capacity is maximum. Our results show that a temperature variable measurement at the fixed working frequency ($T$-scan) allows to determine the overall temperature-dependence of sample heat capacity within an accuracy of 5 $\%$. We also showed that such working frequency might change with pressure, therefore a carefully evaluation through the $f$-scan analysis is recommended to be done at each pressure step to determine absolute values of $C_{S} (T)$.

In our present case, the working frequencies are lower or close to 1 Hz, which demands long-time data acquisition measurement. Our work triggers new investigation about the influence of the sample thickness and relevant parameters in the steady heat equation in order to increase the optimal working frequency.

\section{Acknowledgments}

J. L. J acknowledges to the S\~{a}o Paulo Research Foundation (FAPESP), grants 2019/15912-1 and 2018/08845-3  and CNPq-Universal (431083/2018-5). V. M acknowledges FAPESP (2018/19420-3). H. R and J. L. J acknowledge R. Lortz and M. Zayed for helpful discussion.

\section*{References}

\providecommand{\newblock}{}


\providecommand{\newblock}{}
\begin{thebibliography}{10}
\expandafter\ifx\csname url\endcsname\relax
  \def\url#1{{\tt #1}}\fi
\expandafter\ifx\csname urlprefix\endcsname\relax\def\urlprefix{URL }\fi
\providecommand{\eprint}[2][]{\url{#2}}

\bibitem{Wilhelm_2002}
Wilhelm H and Jaccard D 2002 {\em Journal of Physics: Condensed Matter\/} {\bf
  14} 10683

\bibitem{Lortz2005}
Lortz R, Junod A, Jaccard D, Wang Y, Meingast C, Masui T and Tajima S 2005 {\em
  Journal of Physics: Condensed Matter\/} {\bf 17} 4135

\bibitem{Park2006}
Park T, Ronning F, Yuan H~Q, Salamon M~B, Movshovich R, Sarrao J~L and Thompson
  J~D 2006 {\em Nature\/} {\bf 440} 65

\bibitem{Knebel2011}
Knebel G, Buhot J, Aoki D, Lapertot G, Raymond S, Ressouche E and Flouquet J
  2011 {\em J. Phys. Soc. Jpn\/} {\bf 80} SA001

\bibitem{Larrea2016}
Larrea~J J, Strydom A~M, Martelli V, Prokofiev A, Lorenzer K~A, R\o{}nnow H~M
  and Paschen S 2016 {\em Phys. Rev. B\/} {\bf 93}(12) 125121

\bibitem{Poran2017}
Poran S, Nguyen-Duc T, Auerbach A, Dupuis N, Frydman A and Bourgeois O 2017
  {\em Nature Communications\/} {\bf 8} 14464

\bibitem{Shen2020}
Shen B, Zhang Y, Komijani Y, Nicklas M, Borth R, Wang A, Chen Y, Nie Z, Li R,
  Lu X, Lee H, Smidman M, Steglich F, Coleman P and Yuan H 2020 {\em Nature\/}
  {\bf 579} 51

\bibitem{Lai2017}
Lai H~H, Grefe S~E, Paschen S and Si Q 2018 {\em Proceedings of the National
  Academy of Sciences\/} {\bf 115} 93

\bibitem{Wietek2019}
Wietek A, Corboz P, Wessel S, Normand B, Mila F and Honecker A 2019 {\em Phys.
  Rev. Research\/} {\bf 1}(3) 033038

\bibitem{Gmelin1997}
Gmelin E 1997 {\em Thermochimica Acta\/} {\bf 304-305} 1

\bibitem{Baloga1977}
Baloga J~D and Garland C~W 1977 {\em Review of Scientific Instruments\/} {\bf
  48} 105

\bibitem{Sullivan1968}
Sullivan P~F and Seidel G 1968 {\em Phys. Rev.\/} {\bf 173}(3) 679

\bibitem{Eichler1979}
Eichler A and Gey W 1979 {\em Review of Scientific Instruments\/} {\bf 50} 1445

\bibitem{Gaulin2004}
Gaulin B~D, Lee S~H, Haravifard S, Castellan J~P, Berlinsky A~J, Dabkowska H~A,
  Qiu Y and Copley J~R~D 2004 {\em Phys. Rev. Lett.\/} {\bf 93}(26) 267202

\bibitem{Zayed2017}
Zayed M~E, R{\"u}egg C, Larrea~J J, L{\"a}uchli A~M, Panagopoulos C, Saxena
  S~S, Ellerby M, McMorrow D~F, Str{\"a}ssle T, Klotz S, Hamel G, Sadykov R~A,
  Pomjakushin V, Boehm M, Jim{\'e}nez-Ruiz M, Schneidewind A, Pomjakushina E,
  Stingaciu M, Conder K and R{\o}nnow H~M 2017 {\em Nature Physics\/} {\bf 13}
  962

\bibitem{Choi2002}
Choi E~S, Kang H, Jo Y~J and Kang W 2002 {\em Review of Scientific
  Instruments\/} {\bf 73} 2999

\bibitem{Kageyama2000}
Kageyama H, Onizuka K, Ueda Y, Nohara M, Suzuki H and Takagi H 2000 {\em
  Journal of Experimental and Theoretical Physics\/} {\bf 90} 129

\bibitem{Jorge2005}
Jorge G~A, Stern R, Jaime M, Harrison N, Bon\ifmmode~\check{c}\else \v{c}\fi{}a
  J, El~Shawish S, Batista C~D, Dabkowska H~A and Gaulin B~D 2005 {\em Phys.
  Rev. B\/} {\bf 71}(9) 092403

\bibitem{Berry1963}
Berry R~J 1963  {\bf 41} 946

\bibitem{Dutta2017}
Dutta S, Sankaran K, Moors K, Pourtois G, Van~Elshocht S, Bömmels J,
  Vandervorst W, Tőkei Z and Adelmann C 2017 {\em Journal of Applied
  Physics\/} {\bf 122} 025107

\bibitem{Antonov1984}
Antonov V~E, Belash I~T, Malyshev V~Y and Ponyatovsky E~G 1984 {\em Review of
  Scientific Instruments\/} {\bf 28} 158

\bibitem{Shastry1981}
Shastry B~S and Sutherland B 1981 {\em Physica B+C\/} {\bf 108} 1069

\end{thebibliography}

\end{document}